\documentclass{ws-ijmpb-rev}

\begin{document}

\markboth{N.-C. Yeh et al.}
{Quasiparticle Spectroscopy and High-Field Phase Diagrams ...}

\title{QUASIPARTICLE SPECTROSCOPY AND HIGH-FIELD\\
PHASE DIAGRAMS OF CUPRATE SUPERCONDUCTORS\\
-- AN INVESTIGATION OF COMPETING ORDERS\\
AND QUANTUM CRITICALITY}

\author{N.-C. YEH\footnote{Corresponding author. E-mail: ncyeh@caltech.edu}, C.-T. CHEN, V. S. ZAPF, A. D. BEYER, and C. R. HUGHES}

\address{Department of Physics, California Institute of Technology\\
Pasadena, CA 91125, USA}

\author{M.-S. PARK, K.-H. KIM, and S.-I. LEE}

\address{Pohang University of Science and Technology, Pohang 790-784, Republic of Korea}

\maketitle

\begin{abstract}
We present scanning tunneling spectroscopic and high-field thermodynamic studies of hole- and electron-doped (p- and n-type) cuprate superconductors. Our experimental results are consistent with the notion that the ground state of cuprates is in proximity to a quantum critical point (QCP) that separates a pure superconducting (SC) phase from a phase comprised of coexisting SC and a competing order, and the competing order is likely a spin-density wave (SDW). The effect of applied magnetic field, tunneling current, and disorder on the revelation of competing orders and on the low-energy excitations of the cuprates is discussed.\end{abstract}


\section{Introduction}

There has been emerging consensus that the existence of competing orders\cite{Zhang97,Sachdev03,Kivelson03} in the ground state of cuprate superconductors is likely responsible for a variety of non-universal phenomena such as the pairing symmetry, pseudogap, commensuration of the low-energy spin excitations, and the spectral homogeneity of quasiparticle spectra.\cite{Yeh02a,Yeh01,Yeh02b,Chen02} Among probable competing orders such as the spin-density waves
(SDW),\cite{Demler01,ChenY02} charge-density waves (CDW),\cite{LeeDH02} stripes,\cite{Kivelson03} and the staggered-flux phase,\cite{Kishine01} the dominant competing order and its interplay with superconductivity (SC) remain not well understood. In this work, we report experimental investigation of these issues via quasiparticle spectroscopic and high-field thermodynamic studies. We compare our results with a conjecture that cuprate superconductivity occurs near a quantum critical point (QCP)\cite{Sachdev03,Vojta00} that separates a pure SC phase from a phase with coexisting SC and SDW.\cite{Demler01,ChenY02} The scenario of SDW as the relevant competing order
can be rationalized by the proximity of cuprate SC to the Mott antiferromagnetism, and also by experimental
evidence for spin fluctuations in the SC state of various cuprates.\cite{Wells97,Lake01,Mook02,Yamada03} Possible relevance of SDW in the cuprates to the occurrence of strong quantum fluctuations and the pseudogap phenomenon will be discussed.

\section{Competing Orders and Quantum Criticality}

We consider a conjecture of competing SDW and SC near a non-universal QCP at $\alpha = \alpha _c$,\cite{Demler01} where $\alpha$ is a material-dependent parameter that may represent the doping level, the electronic anisotropy, spin correlation, orbital ordering, or the degree of disorder for a given family of cuprates. As schematically illustrated in Fig. 1(a), in the absence of magnetic field $H$, the ground state consists of a pure SC phase if $\alpha _c < \alpha < \alpha _2$, a pure SDW phase if $\alpha < \alpha _1$, and a SDW/SC coexisting state if $\alpha _1 < \alpha < \alpha _c$. Upon applying magnetic field, delocalized spin fluctuations can be induced due to magnetic scattering from excitons around vortex cores, eventually leading to the occurrence of SDW coexisting with SC for fields satisfying $H^{\ast} (\alpha) < H < H_{c2} (\alpha)$ if the cuprate is sufficiently close to the QCP,\cite{Demler01} and $H_{c2}$ is the upper critical field. In general we expect stronger quantum fluctuations of the SC order parameter in the SDW/SC phase than in the pure SC phase because of excess low-energy excitations associated with the competing SDW. In principle such a difference between coexisting SDW/SC and pure SC phases can be manifested in the field dependence of the thermodynamic properties of the cuprates at $T \to 0$. That is, the proximity of a cuprate to the QCP at $\alpha _c$ can be estimated by determining a characteristic field $H^{\ast}$ using thermodynamic measurements at $T \to 0$, and a smaller magnitude of $(H^{\ast}/H_{c2}^0)$ would indicate a closer proximity to $\alpha _c$ if $\alpha > \alpha _c$, where $H_{c2}^0$ denotes the upper critical field of a given sample at $T = 0$. In contrast, for a cuprate superconductor with $\alpha _1 < \alpha < \alpha _c$, we find $H^{\ast} = 0$ so that SDW coexists with SC even in the absence of external fields, implying gapless SDW excitations (i.e. $\Delta _{SDW} = 0$) and strong excess fluctuations in the SC state. Thus, the thermodynamic quantity $h^{\ast} \equiv (H^{\ast}/H_{c2}^0)$ for a given cuprate is expected to reflect its susceptibility to low-energy excitations and its SC stiffness, which can be confirmed via studies of the quasiparticle spectra taken with a low-temperature scanning tunneling microscope (STM).

\begin{figure}[th]
\centerline{\psfig{file=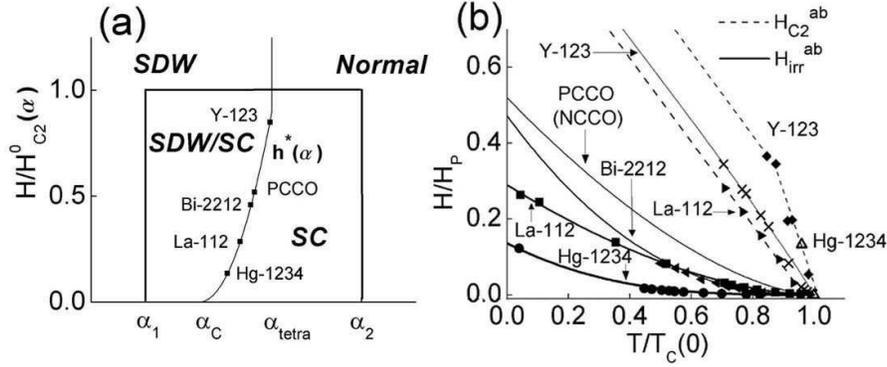,width=12cm}} \vspace*{8pt}
\caption{Phase diagrams of cuprate superconductors based on the
conjecture of competing SDW and SC:$^8$ (a) Reduced
field $(H/H_{c2}^0)$ vs. material parameter ($\alpha$) phase
diagram at $T = 0$. Here $h^{\ast} (\alpha) \equiv
H^{\ast}(\alpha)/H_{c2}^0$ denotes the phase boundary that
separates a pure SC phase from a coexisting SDW/SC phase. The
points along $h^{\ast}$ represent the reduced irreversibility
fields, $H_{irr} (T \to 0)/H_{c2}^0$ obtained from data taken on
Y-123$^{17,18}$, PCCO (NCCO)$^{19,20}$,
Bi-2212$^{21}$, La-112$^{22}$, and Hg-1234.
(b) Comparison of the reduced irreversibility fields (solid lines)
and upper critical fields (dashed lines) vs. reduced temperature
$(T/T_c)$ phase diagram for Hg-1234, La-112, Bi-2212, NCCO, and
Y-123 with $H \parallel$ ab-plane, so that the corresponding
$H_{c2}^0$ is limited by the paramagnetic field $H_p$.}
\end{figure}

\section{Experimental Approach and Results}

To investigate the conjecture outlined above, we employ in this work
measurements of the penetration depth $\lambda (T,H)$, magnetization
$M(T,H)$, and third-harmonic susceptibility $\chi _3 (T,H)$
on different cuprates to determine the irreversibility field
$H_{irr}(T)$ and the upper critical field $H_{c2}(T)$. The degree
of quantum fluctuations in each sample is then estimated by
the ratio $h^{\ast} \equiv (H^{\ast}/H_{c2}^0)$, where the
characteristic field $H^{\ast}$ is defined as
$H^{\ast} \equiv H_{irr}(T \to 0)$. The magnitude of $h^{\ast}$
for different cuprates is determined and compared with the
corresponding quasiparticle spectra.

Specifically, the experiments results reported in this work
consist of studies on the n-type optimally doped infinite-layer
cuprate $\rm La_{0.1}Sr_{0.9}CuO_2$ (La-112, $T_c = 43$ K) and
one-layer $\rm Nd_{1.85}Ce_{0.15}CuO_{4-\delta}$ 
(NCCO, $T_c = 21$ K); and the p-type optimally doped 
$\rm HgBa_2Ca_3Cu_4O_x$ (Hg-1234, $T_c = 125$ K) and 
$\rm YBa_2Cu_3O_{7-\delta}$ (Y-123, $T_c = 93$ K). Results obtained
by other groups on p-type underdoped Y-123 ($T_c = 87$ K),\cite{O'Brien00}
over- and optimally doped $\rm Bi_2Sr_2CaCu_2O_{8+x}$
(Bi-2212, $T_c = 60$ K and 93 K)\cite{Krusin-Elbaum04,Krasnov00};
and n-type optimally doped $\rm Pr_{1.85}Ce_{0.15}CuO_{4-\delta}$
(PCCO, $T_c = 21$ K),\cite{Kleefisch01}
are also included for comparison. The optimally doped samples
of La-112 and Hg-1234 were prepared under high pressures,
with details of the synthesis and characterizations described
elsewhere.\cite{Jung02a,KimMS98,KimMS01} Detailed physical properties of
the optimally doped Y-123 single crystal\cite{Yeh93} and NCCO
epitaxial thin-film\cite{Yeh92} have also been given elsewhere.

The $M(T,H)$ measurements were conducted in lower DC fields
using a Quantum Design SQUID magnetometer at Caltech, and
in high magnetic fields (up to 50 Tesla in a $^3$He refrigerator)
using a compensated coil in the pulsed-field facilities at
the National High Magnetic Field Laboratory (NHMFL) in Los Alamos.
The irreversibility field $H_{irr}(T)$ was identified from the
onset of reversibility in the $M(T,H)$ loops, as exemplified in
the inset of Fig. 2(a) for La-112 and in the main panel of Fig. 2(b)
for Hg-1234. The penetration depths of La-112 and Hg-1234 were
determined in pulsed fields up to 65 Tesla by measuring the
frequency shift $\Delta f$ of a tunnel diode oscillator (TDO)
resonant tank circuit with the sample contained in one of the
component inductors.\cite{Mielke01}
Small changes in the resonant frequency can be related to
changes in the penetration depth $\Delta \lambda$ by
$\Delta \lambda = -\frac{R^2}{r_s}\frac{\Delta f}{f_0}$,
where $R$ is the radius of the coil and $r_s$ is the radius of
the sample.\cite{Mielke01} In our case, $R \sim r_s = 0.7$ mm
and the reference frequency $f_0 \sim 60$ MHz such that
$\Delta f \sim$ (0.16 MHz/$\mu$m)$\Delta \lambda$. Further details
for the pulsed-field measurements of La-112 can be found
in Ref.\cite{Zapf04}. Third-harmonic magnetic susceptibility
$\chi _3 (T,H)$ measurements were also performed on Hg-1234
sample using a 9-Tesla DC magnet and Hall probe techniques.\cite{Reed95}
The $\chi _3 (T,H)$ data measured the non-linear response of the
sample and were therefore sensitive to the occurrence of
phase transformation.\cite{Reed95}

Selected data of these thermodynamic measurements of
La-112 and Hg-1234 are shown in Figs. 2(a)-(b), and a collection
of measured $H_{irr}^{ab}(T)$ and $H_{c2}^{ab}(T)$ curves for various
cuprates are summarized in Fig. 1(b), with the reduced characteristic
fields $(h^{\ast})$ of several representative cuprates explicitly
given in Fig.1(a). We note that the Hg-1234 sample, while having
the highest $T_c$ and upper critical field (estimated at
$H_p \sim H_{c2}^{ab} \sim 500$ Tesla) among all cuprates
shown here, has the lowest reduced irreversibility line
$(H_{irr}^{ab}(T)/H_p)$, where $H_p \equiv \Delta _{SC}^0/(\sqrt{2} \mu _B)$
is the paramagnetic field, and $\Delta _{SC}^0$ denotes the
superconducting gap at $T = 0$. The low irreversibility field is
not only due to the extreme two-dimensionality (2D)
of Hg-1234\cite{KimMS01} that leads to strong thermal
fluctuations at high temperatures, but also due to its
close proximity to the QCP, yielding strong field-induced quantum
fluctuations at low temperatures.

\begin{figure}[th]
\centerline{\psfig{file=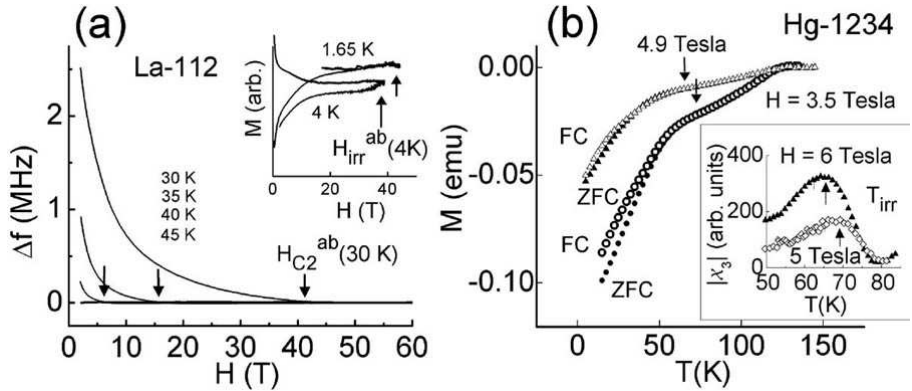,width=12cm}} \vspace*{8pt}
\caption{(a) Main panel: selected data for changes in the resonant
frequency $\Delta f$ of the TDO tank circuit relative to the
normal state of La-112 as a function of $H$ at various $T$. The
estimated $H_{c2}$ values for $H \parallel$ ab-plane,
$H_{c2}^{ab}(T)$, are indicated by arrows. Similar measurements on
a grain-aligned La-112 sample have also been performed (not
shown), which yield $H_{c2}^c(T)$. Inset: $M(T,H)$-vs.-$H$ data on
La-112 for $T = 1.65$ and 4.0 K, where the irreversibility field
$H_{irr}^{ab}(T)$ are indicated by arrows.$^{22}$ (b) Main panel:
representative zero-field-cooled (ZFC) and field-cooled (FC)
$M(T,H)$-vs.-$T$ data of Hg-1234 taken at $H = 3.5$ and 4.9 Tesla,
with the corresponding irreversibility temperature
$T_{irr}^{ab}(H)$ indicated by arrows, using the criterion 
$|M_{ZFC}(T_{irr},H) - M_{FC}(T_{irr},H)|/|M(T \to 0, H \to 0)|
\sim 1.5 \%$. Inset: Representative third-harmonic 
susceptibility$^{28}$ $|\chi _3 (T,H)|$ data for Hg-1234
sample taken at 5 and 6 Tesla, with the corresponding
$T_{irr}^{ab}(H)$ indicated by arrows.}
\end{figure}

To further evaluate our conjecture that cuprates with smaller
$h^{\ast}$ are in closer proximity to a QCP at $\alpha _c$ and
are therefore associated with a smaller SDW gap $\Delta _{SDW}$ and
stronger SC fluctuations, we examine the
SC energy gap $\Delta _{SC} (T)$ and the quasiparticle
low-energy excitations of different cuprates. In Fig. 3(a),
we compare the $\Delta _{SC}(T)$ data of La-112, taken with our
low-temperature scanning tunneling microscope (STM), with those of
Bi-2212 and PCCO obtained from intrinsic tunnel junctions\cite{Krasnov00}
and grain-boundary junctions,\cite{Kleefisch01} respectively. Noting
that the $h^{\ast}$ values are $\sim 0.53$ for PCCO (NCCO), $\sim 0.45$
for Bi-2212, and $\sim 0.24$ for La-112, we find that the rate of
decrease in $\Delta _{SC}$ with $T$ also follows the same trend.
These differences in $\Delta _{SC}(T)$ cannot be attributed
to different pairing symmetries, because La-112 is consistent with
$s$-wave pairing symmetry,\cite{Chen02} NCCO (PCCO) can exhibit
$s$-wave\cite{Alff99} or $d$-wave pairing,\cite{Tsuei00} depending on
both the cation and oxygen doping levels,\cite{Skinta02,Biswas02} 
and Bi-2212 is $d$-wave pairing.\cite{Renner98,Pan00,Hudson01}
Therefore, the sharp contrast in the $\Delta _{SC}(T)$ data between
La-112 and NCCO (PCCO) suggests that the proximity to the QCP
at $\alpha _c$ plays an important role in determining the low-energy
excitations of the cuprates. These experimental findings have been
further corroborated by recent $t$-$t^{\prime}$-$U$-$V$ model
calculations\cite{ChenHY04} for optimally doped p-type cuprates, which
demonstrate that the competing order SDW can appear with increasing 
temperature even though $\alpha > \alpha _c$ at $T = 0$.  

Another experimental confirmation for our conjecture can be found
in the quasiparticle tunneling spectra exemplified in Fig. 3(b).
We find that excess sub-gap spectral weight exists in La-112,
although the quasiparticle spectra of La-112 are momentum-independent
and its response to quantum impurities is consistent with
$s$-wave pairing.\cite{Yeh02b,Chen02,Yeh03a} The excess sub-gap spectral
weight relative to the BCS prediction implies excess low-energy
excitations that cannot be reconciled with simple quasiparticle
excitations from a pure SC state, and is therefore strongly
suggestive of the presence of a competing order with an additional
channel of low-energy excitations. In contrast, substantial
details of the quasiparticle spectrum on optimally doped Y-123
(for quasiparticle energies $|E|$ up to $>\sim \Delta _{SC}$)
can be explained by the generalized BTK theory,\cite{Yeh01,Wei98}
implying that the spectral contribution due to the competing order
is insignificant up to $|E| \sim \Delta _{SC}$. This finding is
in agreement with a much larger $h^{\ast}$ value for Y-123 than for 
La-112, and therefore much weaker quantum fluctuations in the former.

\begin{figure}[th]
\centerline{\psfig{file=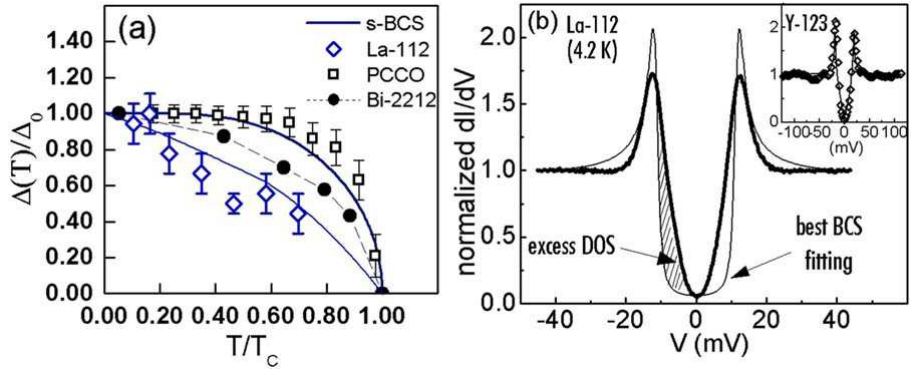,width=12cm}}
\vspace*{8pt}\caption{(a) Comparison of the temperature ($T$)
dependence of the normalized superconducting energy gap, $\lbrack
\Delta _{SC}(T) / \Delta _{SC}^0 \rbrack$, for NCCO
(PCCO),$^{20}$ Bi-2212,$^{23}$ and La-112. (b)
Main panel: comparison of the normalized quasiparticle spectrum of
La-112 (points), obtained using a STM at $T = 4.2$ K, with the
best BCS fitting (solid line), showing excess sub-gap and reduced
post-gap spectral weight. Inset: comparison of the normalized
c-axis quasiparticle tunneling spectrum (points) of Y-123, taken
using a STM at 4.2 K, with the generalized BTK
fitting$^{5,38}$ for $d_{x^2-y^2}$-wave superconductors
(solid line). We note quality agreement between the generalized
BTK fitting and data up to $|E| \sim \Delta _{SC}$.}
\end{figure}

A further verification for the closer proximity of La-112 to
the QCP than Y-123 is manifested in Figs. 4(a)-(b), where the
quasiparticle tunneling spectra of La-112 are found to be
dependent on the tunneling current $I$. For lower tunneling currents,
we find that the coherence peaks at $E = \pm \Delta _{SC}$
are systematically suppressed by increasing $I$ without changing
the $\Delta _{SC}$ value. The coherence peaks eventually vanish
while additional features at higher energies begin to emerge.
Finally, in the large current limit, pseudogap-like features
appear at $|E| \equiv \Delta _{PG} > \Delta _{SC}$, as illustrated
in Fig. 4(a). More detailed evolution of the superconducting and
pseudogap features with $I$ is summarized in Fig. 4(b). We also
notice that the magnitude of $\Delta _{SC}$ determined under smaller
tunneling currents is highly spatially homogeneous, whereas
the $\Delta _{PG}$ value appears to vary significantly
from one location to another. We suggest two effects associated
with increasing tunneling current $I$: First, the localized high
current density under the STM tip can effectively suppress the SC
order parameter and therefore reduce $\alpha$. Second, the 
magnetic field induced by the localized high current density 
can also assist the evolution from an initial SC phase with 
$\alpha >\sim \alpha _c$ into the coexisting SDW/SC state with 
increasing $I$. The current-induced SDW can be manifested in 
the DC tunneling spectrum through coupling of the SDW order 
parameter to disorder.\cite{Polkovnikov02}
Hence, the resulting quasiparticle spectrum under large tunneling
currents contains convoluted spectral information of SDW, SC,
and the disorder potential, thus it can be spatially
inhomogeneous. In contrast, for optimally doped Y-123, no
noticeable spectral variation was found with tunneling
currents in the same range as that used for studying La-112. This
finding is again consistent with our conjecture that Y-123
is farther from the QCP than La-112, so that it is more difficult
to induce SDW in Y-123.

\begin{figure}[th]
\centerline{\psfig{file=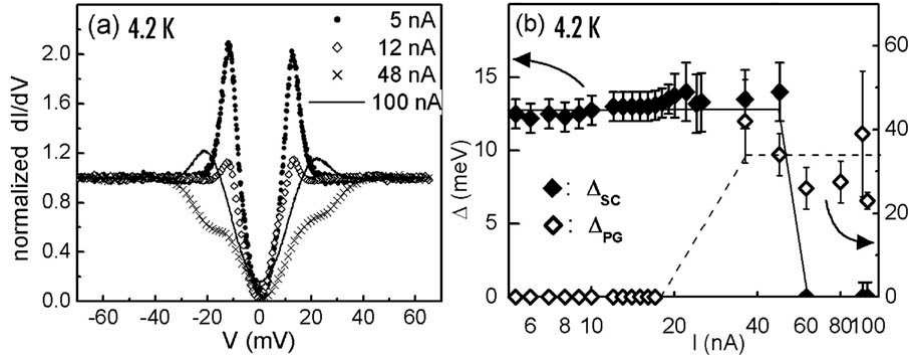,width=12cm}}
\vspace*{8pt}\caption{(a) Evolution of the normalized
quasiparticle tunneling spectrum of La-112 with increasing
tunneling current $I$, taken at 4.2 K. (b) Evolution of $\Delta
_{SC}$ and $\Delta _{PG}$ of La-112 with $I$, as determined from
the quasiparticle tunneling spectra.}
\end{figure}

\section{Disorder Effect on Quasiparticle Spectra and Pseudogap}

Next, we investigate the effect of disorder on the scenario
depicted in Fig. 1(a). Generally speaking, disorder reduces
the SC stiffness and tends to shift $\alpha$ closer
to $\alpha _c$ if initially $\alpha > \alpha _c$.\cite{Vojta00} 
Therefore one can envision spatially varying $\alpha$ values in 
a sample if the disorder potential is spatially inhomogeneous. In
particular, for strongly 2D cuprates like Bi-2212 and Hg-1234,
disorder can help pin the fluctuating SDW locally,\cite{Yeh03b}
so that regions with the disorder-pinned SDW can coexist with SC,
as schematically illustrated
in Fig. 5(a). These randomly distributed regions of pinned SDW
are scattering sites for quasiparticles at $T < T_c$
and for normal carriers at $T > T_c$, provided that the SDW
persists above $T_c$. We have performed numerical calculations
for the quasiparticle local density of states (LDOS) of a
2D $d$-wave superconductor to examine these related issues.\cite{Chen03}
We have considered two scenarios: one assumes that
the ground state of the 2D $d$-wave cuprate is a pure SC
with random point defects, and the other assumes that randomly
pinned SDW regions coexist with SC at $T \ll T_c$. Using
the Green's function techniques detailed in Ref.\cite{Chen03},
we find that the quasiparticle interference spectra for
SC coexisting with a disorder-pinned SDW differ fundamentally
from those due to pure SC with random point disorder.
A representative real-space map of the quasiparticle LDOS, for
a 2D $d_{x^2-y^2}$-wave superconductor with 24 randomly
distributed pinned SDW regions in an area of $(400 \times 400)$
unit cells, is shown in Fig. 5(b). The Fourier transformation
(FT) of the LDOS is illustrated in Fig. 5(c) for a pure SC
with random point disorder at $T = 0$, in Fig. 5(d) for the
FT-LDOS of pinned SDW at $T = 0$, and in Fig. 5(e) for the
FT-LDOS of pinned SDW at $T = T_c$.
We note that the superposition of the FT-LDOS in Figs. 5(c)
and 5(d) is consistent with experimental observation on a slightly
underdoped Bi-2212 at $T \ll T_c$,\cite{Hoffman02,McElroy03}
whereas the FT-LDOS in Fig. 5(e) is consistent with experimental
observation on a similar sample at $T > \sim T_c$.\cite{Vershinin04}
These findings suggest that a disorder-pinned SDW coexists with
SC in Bi-2212 at low temperatures, and that only the SDW persists
above $T_c$.\cite{Chen03} These studies therefore demonstrate
the significant role of competing orders in determining
the physical properties of cuprate superconductors.
Moreover, disorder-pinned collective modes such as SDW
can naturally account for the strong spatial inhomogeneity
observed in the quasiparticle spectra of Bi-2212,\cite{Pan01,Lang02}
and are also likely responsible for the pseudogap phenomenon\cite{Timusk99}
above $T_c$ in under- and optimally doped p-type cuprates.

\begin{figure}[th]
\centerline{\psfig{file=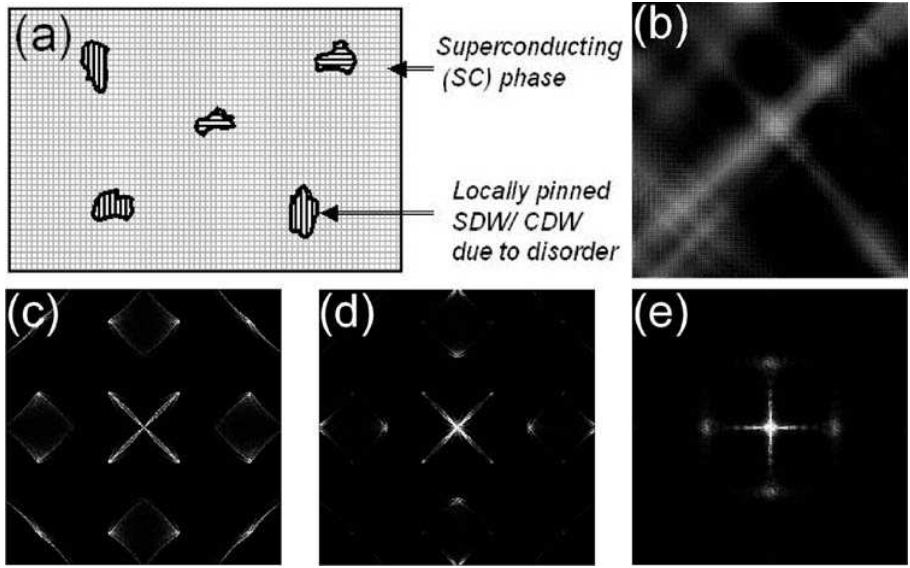,width=12cm}}
\vspace*{8pt}\caption{(a) Schematic illustration of randomly
distributed disorder-pinned SDW regions in a 2D superconductor.
(b) Quasiparticle LDOS for the scenario depicted in (a), for a
$d_{x^2-y^2}$-wave superconductor at $T = 0$, with a SC gap
$\Delta _{SC} = 40$ meV and the quasiparticle energy $E = 24$ meV.
(c) Quasiparticle FT-LDOS in the first Brillouin zone of a 2D
$d_{x^2-y^2}$-wave superconductor with random point defects at $T
= 0$. (d) The FT-LDOS of Part (b) in the first Brillouin zone at
$T = 0$. (e) The FT-LDOS of Part (b) at $T = T_c$. More details
are given in Ref. ${41}$.}
\end{figure}

\section{Discussion}

The experimental results presented in the previous sections
are generally consistent with the notion that significant
quantum fluctuations can be induced by a magnetic field in
all cuprate superconductors, and that the degree of excess
low-energy excitations can be correlated with the
proximity of the cuprates to a QCP. However, to further our
understanding, systematic studies of more cuprates
will be necessary. In particular, neutron scattering studies
will be important for determining the SDW gaps
of different cuprates. It is also imperative to examine the
correlation between the low-temperature high-field phase diagram
and the low-energy excitations of different cuprates through
quasiparticle spectroscopic studies. For instance, quasiparticle
tunneling spectroscopic studies of the highly 2D Hg-1234 can
substantiate our conjecture if the following can be verified:
(1) $\Delta _{SC}(T)$ decreases more rapidly
with $T$ than other p-type cuprates with larger values of $h^{\ast}$;
(2) the quasiparticle DOS exhibits strong spatial variation below
$T_c$, similar to Bi-2212; (3) excess sub-gap DOS than the
BTK prediction exists because of the relatively smaller $\Delta _{SDW}$
that provides an additional channel of low-energy excitations; and
(4) strong quasiparticle spectral dependence on the tunneling
current. Similarly, determination of the characteristic
field $h^{\ast}$ as a function of impurity concentration can provide
further verification for our conjecture that $h^{\ast}$
varies with increasing disorder.

Despite consensus for the existence of competing orders in the ground
state of the cuprates, whether SDW is the dominant competing order is
yet to be further verified. For instance, we note that
certain experimental consequences (such as the quasiparticle spectra)
due to a disorder-pinned SDW cannot be trivially distinguished from
a disorder-pinned CDW in tetragonal cuprates. Therefore neutron
scattering studies will be necessary to distinguish between these two
competing orders. As for the staggered flux phase, unit-cell 
doubling features for the quasiparticle LDOS along the CuO$_2$ 
bonds must be demonstrated around a vortex core if the staggered 
flux phase is a relevant competing order.\cite{Kishine01} 
Finally, the most important issue remaining to be addressed 
in the studies of competing orders and quantum
criticality is to investigate possible correlation between
the existence of competing orders and the occurrence of
cuprates superconductivity.

\section{Summary}

In summary, we have investigated the quasiparticle tunneling spectra
and the thermodynamic high-field phase diagrams of various
electron- and hole-doped cuprate superconductors. The experimental
data reveal significant magnetic field-induced quantum fluctuations
in all cuprate superconductors, and the degree of
quantum fluctuations appears to correlate well with the magnitude
of excess low-energy excitations as the result of competing
orders in the ground state. Moreover, our experimental
results support the notion that the ground state of cuprates
is in proximity to a QCP separating pure SC from coexisting SC/SDW,
and our theoretical analysis further suggests that disorder-pinned
fluctuating SDW can have significant effect on the LDOS of highly
2D cuprates. Additionally, the persistence of disorder-pinned SDW 
above $T_c$ may be accountable for the pseudogap phenomenon observed
in the spectroscopic studies of highly 2D hole-doped cuprates.

\section*{Acknowledgments}

The work at Caltech is supported by the National Science Foundation
through Grants \#DMR-0405088 and \#DMR-0103045, and at the Pohang University
by the Ministry of Science and Technology of Korea through the 
Creative Research Initiative Program. The pulsed-field experiments were
performed at the National High Magnetic Field Laboratory facilities in the
Los Alamos National Laboratory under the support of the National Science
Foundation.

\end{document}